\documentclass[11pt]{article}
\usepackage{moriond,epsfig}

\def\Journal#1#2#3#4{{#1} {\bf #2}, #3 (#4)}

\def\PLB{{\em Phys. Lett.}  B}

\def\PRD{{\em Phys. Rev.} D}

\def\ra{\rightarrow}
\def\be{\begin{equation}}
\def\ee{\end{equation}}
\def\bea{\begin{eqnarray}}
\def\eea{\end{eqnarray}}

\def\ipb{pb^{-1}}
\def\psits{\psi(3770)}
\def\gev{GeV}
\def\mev{MeV}
\def\ebeam{E_{\mathrm{beam}}}
\def\d{D}
\def\b{B}
\def\ds{D_{s}}
\def\bt{T}
\def\vcd{V_{cd}}
\def\vcs{V_{cs}}
\def\vcb{V_{cb}}
\def\vub{V_{ub}}
\def\vtd{V_{td}}
\def\vts{V_{ts}}

\def\dtomn{D^{+} \ra \mu^{+} \nu}

\def\dotoken{D^{0} \ra K^- e^{+} \nu}
\def\dotopien{D^{0} \ra \pi^- e^{+} \nu}
\def\dotoksen{D^{0} \ra K^{*-} e^{+} \nu}
\def\dotoksentwo{D^{0} \ra K^{*-} (K^- \pi^0) e^{+} \nu}
\def\dotoksenthree{D^{0} \ra K^{*-} (K_S^0 \pi^-) e^{+} \nu}
\def\dotorhoen{D^{0} \ra \rho^- e^{+} \nu}

\def\dtoken{D^{+} \ra \overline{K}^0 e^{+} \nu}
\def\dtopien{D^{+} \ra \pi^0 e^{+} \nu}
\def\dtoksen{D^{+} \ra \overline{K}^{*0} e^{+} \nu}
\def\dtoksentwo{D^{+} \ra \overline{K}^{*0} (K^- \pi^+) e^{+} \nu}
\def\dtorhoen{D^{+} \ra \rho^0 e^{+} \nu}
\def\dtorhoentwo{D^{+} \ra \rho^0 (\pi^+ \pi^-) e^{+} \nu}
\def\dtowen{D^{+} \ra \omega e^{+} \nu}
\def\dtowentwo{D^{+} \ra \omega (\pi^+ \pi^- \pi^0) e^{+} \nu}

\begin{document}
\vspace*{4cm}
\title{RECENT RESULTS FROM CLEO}

\author{DORIS YANGSOO KIM \\
        Representing the CLEO-c Collaboration}

\address{Loomis Lab of Physics, 1110 W Green St,\\
Urbana, IL. 61801, USA}

\maketitle\abstracts{
We report the initial results from CLEO-c based on the
55.8 $\ipb$ data obtained at the $\psits$ resonance last year. 
We give a concise summary of the various CLEO-c analyses on leptonic and
semileptonic decays of neutral and charged \d mesons.
The data used for this report is the
first part of the $\psits$ data sets to be collected during next several years
using the Cornell Electron Storage Ring. Most of the results shown here have
already better or the same level of precision compared to the 2004 world
averages compiled by the Particle Data Group.~\cite{pdg2004}}

\section{CESR-c and CLEO-c}
The venerable Cornell Electron Storage Ring (CESR) had been running at the $\b$
physics energy region during last decade. Recently, a decision was made to
run CESR at lower energies to pursue next generation precision physics
in other interesting fields including the charm sector.
To accomplish this task, twelve superconducting wigglers have been 
added to the Storage Ring. The wigglers increase the synchrotron radiation of
electron and positron beams, damping the beams. As the result, the 
luminosity at low beam energies becomes increased. The modified
Storage Ring can handle beam energies between 1.5 and 5.6 $\gev$. The
current run plan includes CESR operations at the $\psi^{\prime}$,
$\psits$ and $J/\psi$ resonances and the $\ds^+\ds^-$ pair threshold area.

The CLEO-c detector is almost the same as the CLEO III detector. Most of the
CLEO III detector components are retained. The old
silicon vertex detector of CLEO III was replaced by a new inner drift chamber
with six layers. The magnetic field was decreased from 1.5 $\bt$ to
1.0 $\bt$ to accommodate lower beam energy accelerator operations. 

\subsection{Impacts of the CLEO-c Measurements}

The leptonic and semileptonic decay kinematics of $\d$ mesons is 
computed from first principles using the Cabibbo-Kobayashi-Maskawa
(CKM) quark mixing matrix elements. The hadronic complications are
contained in a few functional forms. 
In the leptonic decays, the strong interactions are represented by
decay constants $\mathrm{f_{D^+}}$ and $\mathrm{f_{D_{S}}}$ for $\d^{+}$ and
$\ds^{+}$ decays, respectively, which can be calculated by Lattice QCD.
In the semileptonic decays, the strong interactions are represented by form
factors, which are calculable by various methods. The precision measurement
of these parameters will validate and enhance Lattice QCD calculations and
test theoretical form factor models, which will improve $\b$ meson decay
constant and form factor calculations. Subsequently, all these improvements
will be used to test unitarity of the CKM matrix. 

The large size of samples with fairly clean background conditions
expected from CESR-c operations implies 
excellent opportunities to improve branching fraction measurements of
many important decay modes of $\d$ mesons. Hadronic decays of $\d$ mesons
provide information for strong interaction phases essential for CP violation
studies. The initial states of $\d$ mesons created at CESR-c are coherent,
which generate an ideal environment to study mixing and CP violation at
the charm sector. 

One of the most important goals of the precision flavor physics is to measure
all the CKM matrix elements and associated phases, to over-constrain the unitary
triangles. CLEO-c will contribute to this effort by reducing uncertainties of
$\vcd$ and $\vcs$ down to the 1.7\% level by exploiting $\d$ meson decays.
Together with the coming results from Lattice QCD, $\b$ factories and
$\mathrm{p\overline{p}}$ collider experiments, the precision of the related
CKM matrix elements in the beauty and top sectors will be greatly improved. The
uncertainty on $\vcb$ is expected to be reduced to the 3\% level and 
the uncertainties
on $\vub$, $\vtd$ and $\vts$ are expected to be reduced to 
the 5\% level.~\cite{yellowbook}  
        
\subsection{Analysis Techniques at the $\psits$ Resonance}

When the electron and positron beams of CESR collide at the $\psi(3770)$
resonance, the resulting $\d\overline{\d}$ pairs are created at threshold
energy. There are no extra fragmentation particles other than the $\d$ mesons
at the production vertex, which provides a simple geometry to reconstruct
the event. The combinatoric backgrounds coming from wrong assignment of decay
tracks are small. Reconstruction of unseen neutrinos in leptonic and
semileptonic decays of $\d$ mesons becomes quite clean and straightforward.
To identify the observed events as $\d\overline{\d}$ pairs, we first
reconstruct a $\d$ meson fully as ``the tag," then we analyze the decay of the
second $\d$ in the same event to extract exclusive or inclusive properties.
The tagging efficiency of $\d\overline{\d}$ pair events is high, estimated as
$\approx$ 25\% of all the $\d$ mesons produced. (Charge conjugate modes are
implied in this report.) 

\subsection{Tagged $\d$ Samples from CLEO-c}

From December 2003 to March 2004, we collected the first 55.8 $\ipb$ of CLEO-c
data at the $\psits$ resonance with 6 wigglers, generating 360,000
$\d\overline{\d}$ pairs. All the studies except the last topic on the
CP violation shown in this report are based on this data set.
After applying basic selection cuts, we used beam energy information and
4-momentum conservation condition to finalize tagged $\d$ meson samples. The
variables used for this stage of selection and for estimating the number
of signals are the difference in the energy ($\Delta E = E(D) - \ebeam$)
and the beam energy constrained mass
($M_{BC} = \sqrt{\ebeam^2 - | p(D) |^2}$).
For example, to measure exclusive semileptonic branching fractions of
$\d^0$ ($\d^+$) mesons, we selected $\approx$ 60,000 (32,000) decays using
eight (six) hadronic decay modes of $\overline{\d}^0$ ($\d^-$) as tags. 

\section{Leptonic Decays of $\d$ Mesons: $\dtomn$}

We published a clean observation of the $\dtomn$ decays last 
year.~\cite{dtomnpaper} To select pure $\dtomn$ signals, we required
a $\d^{-}$ hadronic tag in the system as the other $\d$ meson and an additional
charged track as the $\mu^{+}$ from the signal $\d$ meson. We limited the
energy of most energetic extra shower to be less than 250 $\mev$.
Since there should be one unseen neutrino in the system, the genuine signal
events have to generate a peak at 0 when we plot distributions
of ``missing mass squared",
($\mathrm{MM}^2 = 
 (\ebeam - E_{\mu^+})^2 - (-\vec{p}_{D^-} - \vec{p}_{\mu^+})^2$).
 
\begin{figure}[p]
\begin{center}
\psfig{figure=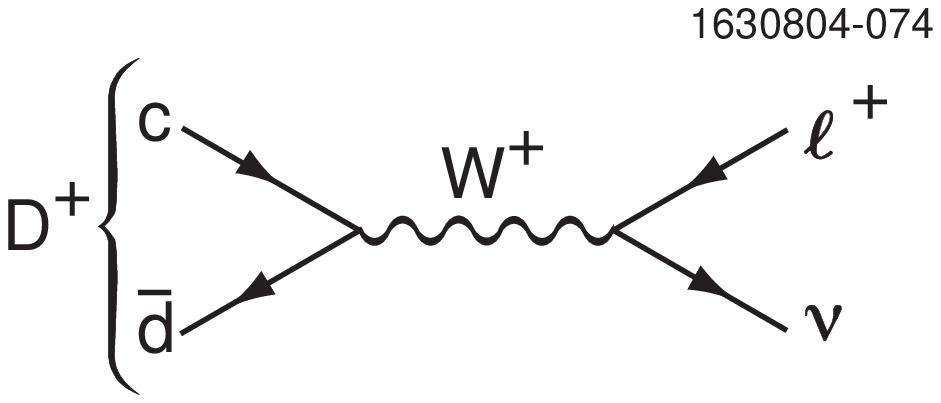,height=1.8in}
\psfig{figure=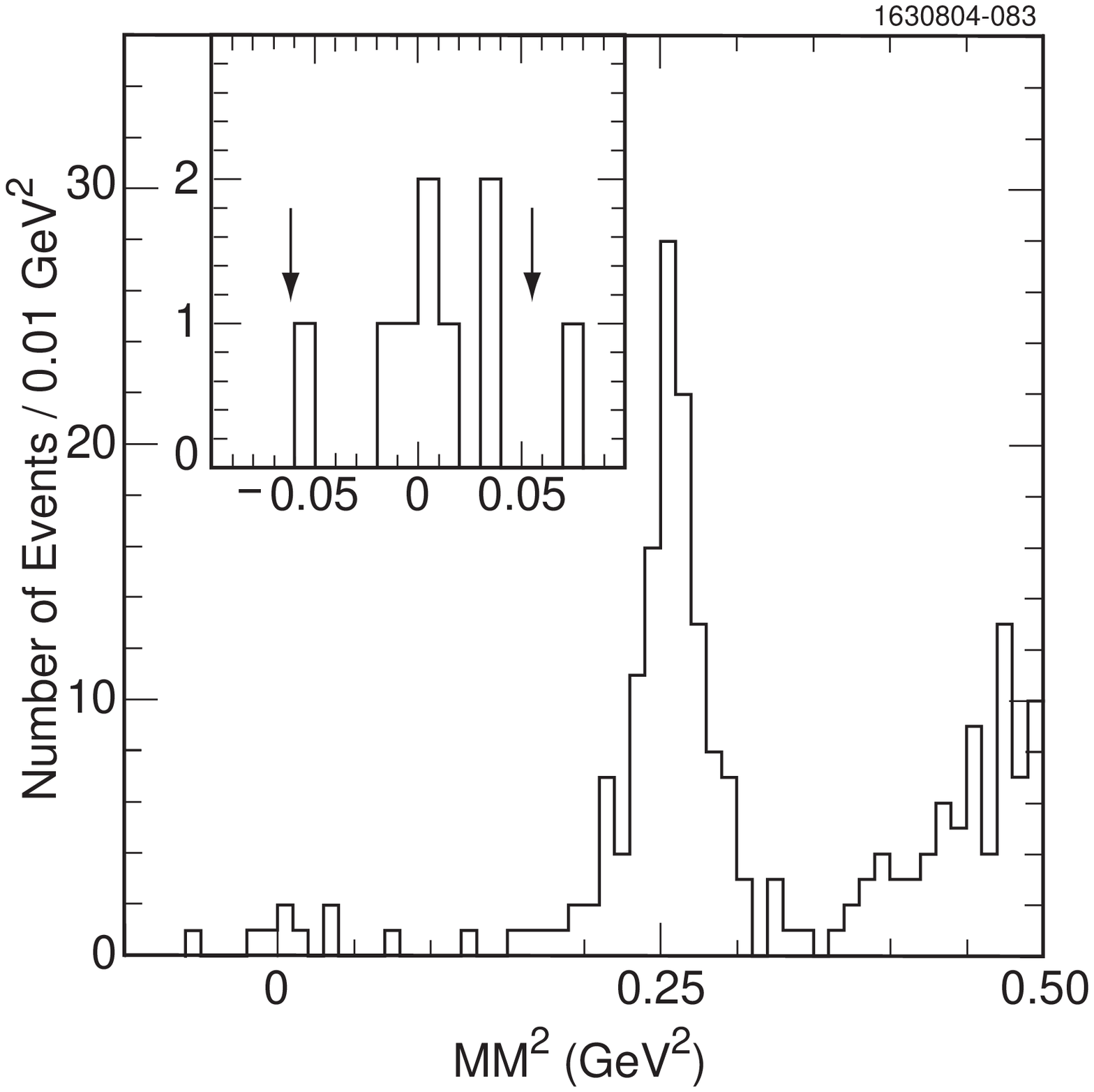,height=2.5in}
\end{center}
\caption{The left plot is the Feynman diagram of the $\dtomn$ decays.
The right plot is the distribution of missing mass squared of the
$\dtomn$ candidates.
The inset plot shows the details in the signal area with the arrows indicating
selection cuts. The larger peak at 0.25 $\gev^2$ represents backgrounds coming
from $D^+ \ra \overline{K}^0 \pi^+$ decays.
\label{fig:dtomn} }
\end{figure}

Figure~\ref{fig:dtomn} shows the distribution of
missing mass squared from our data. From 28651 $\d^-$ tagged events,
we found eight signal candidates. The size of the backgrounds at the signal
area was estimated as one event.  The larger peak at 0.25 $\gev^2$ represents
backgrounds coming from $D^+ \ra \overline{K}^0 \pi^+$ decays. A $K_L$ may
escape the detector without depositing energy in the Electromagnetic calorimeter,
faking a neutrino signal.  We measured
branching fraction of our signal mode as $B(\dtomn) = (3.5 \pm 1.4 \pm 0.6)
\times 10^{-4}$. From this number, we extracted the decay constant
$\mathrm{f_{D^+}}$ using the following equation,

\be
 B(\dtomn)/\tau_{D^+} = \frac{G^2_F}{8\pi} f^2_{D^+} m^2_{\mu}
          M_{D^+} \left( 1 - \frac{m^2_{\mu}}{M^2_{D^+}} \right) |\vcd|^2
\ee 
Since the uncertainty of $\vcd$ is 1.1 \% from the 3 generation unitarity
constraint and the uncertainty of $\tau_{D^+}$ from experimental measurements
is only 0.3 \%, we obtained
$\mathrm{f_{D^+}} = (202 \pm 41 \pm 17)\ \mev$, which is far more accurate than
the recent measurement by BES II, $ (371 ^{+129}_{-119} \pm 25)\ \mev$, based
on three observed events.~\cite{bes}

\section{Exclusive Semileptonic Branching Fractions of $\d$ Mesons}

\begin{figure}[p]
\begin{center}
\psfig{figure=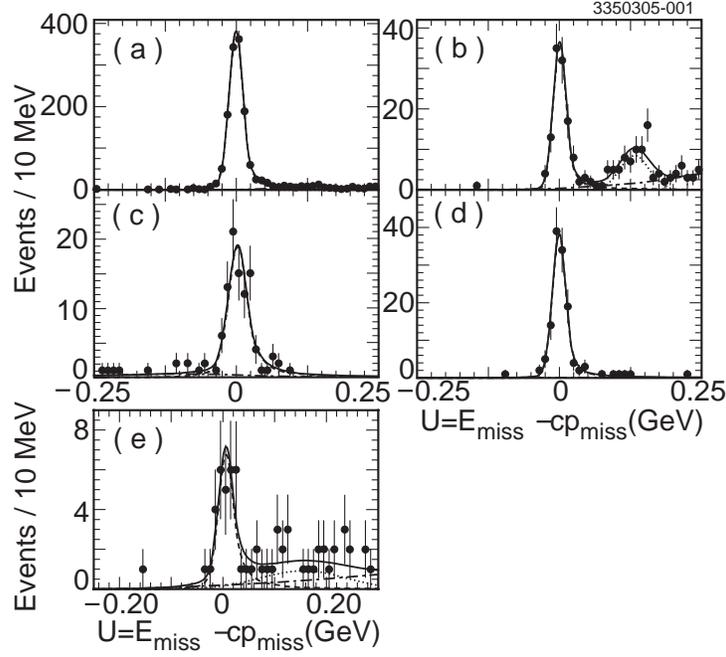,height=3.5in}
\end{center}
\caption{Plots of $U = E_{miss} - |P_{miss}|$ for the study
of exclusive branching fractions of $\d^0$ decays. Plots (a), (c) and (d)
represent Cabibbo favored while Plots (b) and (e) represent Cabibbo
suppressed decay modes. The smaller peaks next to the signal peaks in the 
Cabibbo suppressed mode plots represent backgrounds coming from
the corresponding Cabibbo favored modes. (a) $\dotoken$, (b) $\dotopien$,
(c) $\dotoksen$ ($K^{*-} \ra K^- \pi^0$),
(d) $\dotoksen$ ($K^{*-} \ra K_S^0 \pi^-$),
(e) $\dotorhoen$ ($\rho^- \ra \pi^- \pi^0$). We observed $\dotorhoen$ decays for
the first time in the world. 
\label{fig:exd0signal} }
\end{figure}

\begin{figure}[p]
\begin{center}
\psfig{figure=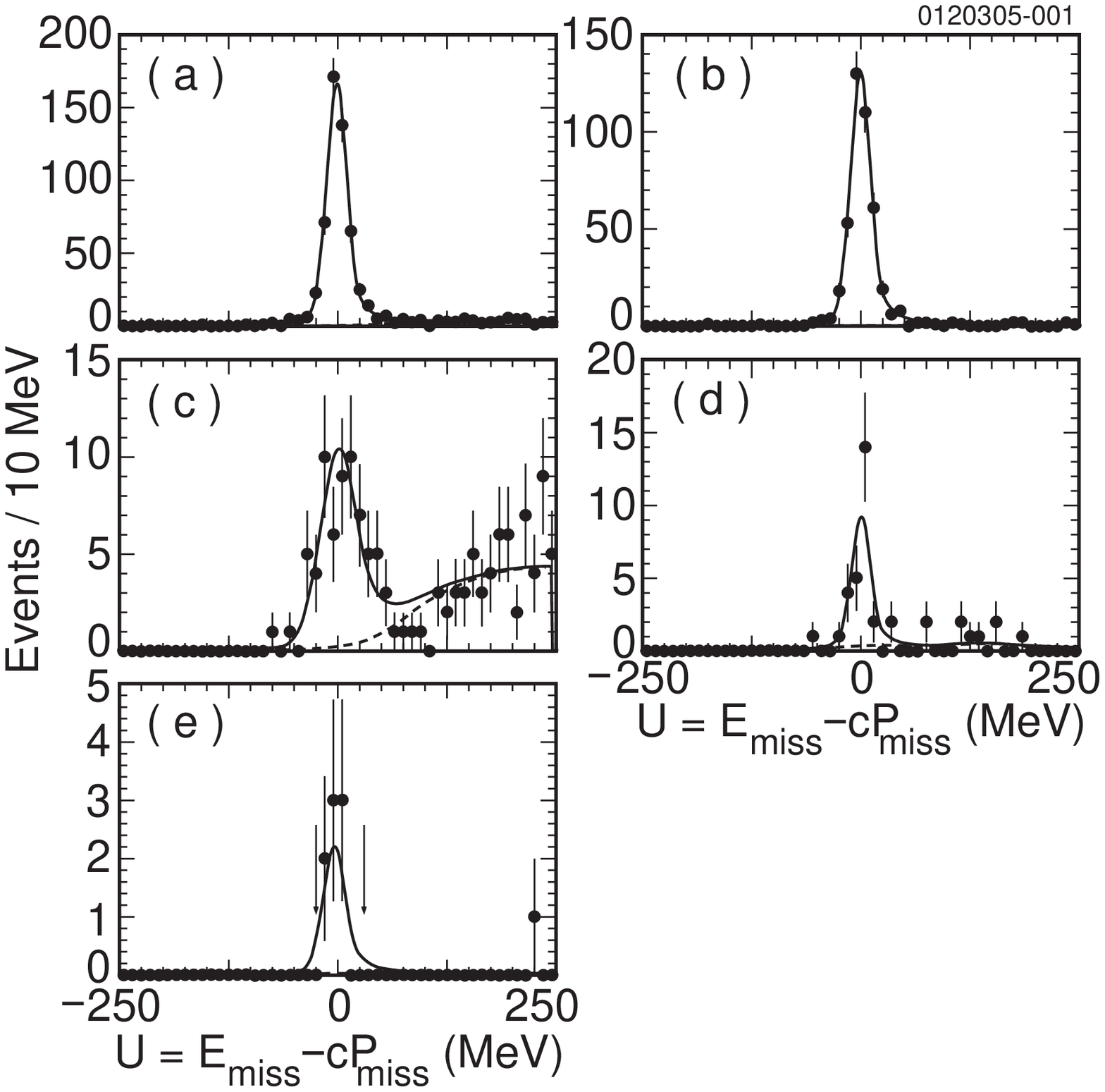,height=3.5in}
\end{center}
\caption{Plots of $U = E_{miss} - |P_{miss}|$ for the study
of exclusive branching fractions of $\d^+$ decays. The left column plots 
of the first two rows represent Cabibbo favored decay modes. All the other
plots represent Cabibbo suppressed decay modes. (a) $\dtoken$, (b) $\dtopien$,
(c) $\dtoksen$ ($\overline{K}^{*0} \ra K^- \pi^+$),
(d) $\dtorhoen$ ($\rho^0 \ra \pi^+ \pi^-$),
(e) $\dtowen$ ($\omega \ra \pi^+ \pi^- \pi^0 $). We observed $\dtowen$ decays
for the first time in the world.
\label{fig:exdpsignal} }
\end{figure}

To select semileptonic decays of $\d^0$ and $\d^+$ mesons, we first required
one fully reconstructed hadronic $\d$ tag as the other $\d$ meson. From the
remaining tracks and showers in the same event, we reconstructed a semileptonic
signal candidate. We used a fit variable $U = E_{miss} - |P_{miss}|$ to 
estimate the number of signal events. Since there should be one unseen neutrino
in the system, this variable produces a prominent signal peak at 0.
We studied four decay channels of $\d^0$ and five decay channels of $\d^+$. 
Two of these decay channels were observed for the first time:
$\dotorhoen$ and $\dtowen$. 
The $U$ distribution of each channel is shown in Figures~\ref{fig:exd0signal}
and \ref{fig:exdpsignal} and the results are summarized in
Table~\ref{tab:ex}.~\cite{exd0report,exdreport}
The uncertainties of our measurements are better or at the
same level of the 2004 world averages compiled by the Particle Data
Group (PDG).~\cite{pdg2004}

\begin{table}[p]
\caption{The exclusive branching fraction measurements of
$\d^0$ and $\d^+$ meson decays based on the first 55.8 $\ipb$ data from CLEO-c
at the $\psits$ resonance. The branching fractions for $\dotoksen$ and 
$\dtoksen$ are reduced by 2.4\% to subtract the non-resonant s-wave
contribution.\label{tab:ex} }
\begin{center}
\begin{tabular}{|l|c|c|}
\hline
Decay Mode  &   $B$ (\%) (here)                &  $B$ (\%) (PDG 2004) \\
\hline
$\dotoken$       &  $3.44 \pm 0.10 \pm 0.10$    & $3.58 \pm 0.18$ \\
$\dotopien$      &  $0.262 \pm 0.025 \pm 0.008$ & $0.36 \pm 0.06$ \\
$\dotoksentwo$   &  $2.11 \pm 0.23 \pm 0.10$    &                 \\
$\dotoksenthree$ &  $2.19 \pm 0.20 \pm 0.11$    &                 \\
$\dotoksen$      &  $2.16 \pm 0.15 \pm 0.08$    & $2.15 \pm 0.35$ \\
$\dotorhoen$     &  $0.194 \pm 0.039 \pm 0.013$ &                 \\
                 &                              &                 \\
$\dtoken$       &  $8.71 \pm 0.38 \pm 0.37$     & $6.7  \pm 0.9$ \\
$\dtopien$      &  $0.44 \pm 0.06 \pm 0.03$     & $0.31 \pm 0.15$ \\
$\dtoksentwo$   &  $5.56 \pm 0.27 \pm 0.23$     & $5.5  \pm 0.7$ \\
$\dtorhoentwo$  &  $0.21 \pm 0.04 \pm 0.01$     & $0.25 \pm 0.10$ \\
$\dtowentwo$    &  $0.16 ^{+0.07}_{-0.06} \pm 0.01$    &          \\
\hline
\end{tabular}
\end{center}
\end{table}

\section{Studies on Inclusive Decays of $\d$ Mesons}

The uncertainties of the current world average of the inclusive semileptonic
branching fractions are 0.28 \% and 1.9 \% for $\d^0 \ra X e^+ \nu$ and 
$\d^+ \ra X e^+ \nu $ decays, respectively. The preliminary study on the first
55.8 $\ipb$ data showed that our statistical uncertainties are expected at the
level of 0.2 \% and 0.3 \% for these decays. We required one hadronic $\d$ tag
as the other $\d$ meson, and one good electron identified by $dE/dx$, RICH
and Electromagnetic calorimeter information. We used the sign relation 
between the signal and the tag $\d$'s to subtract wrong sign events as
backgrounds. (For $\d^0$, the signal charge is defined by the charge of its
decay particle, $K$.)

\section{CP Violation search in $\d^0 \ra K_s \pi^+ \pi^-$ Decays
Based on the CLEO II.V data}

\begin{figure}[p]
\begin{center}
\psfig{figure=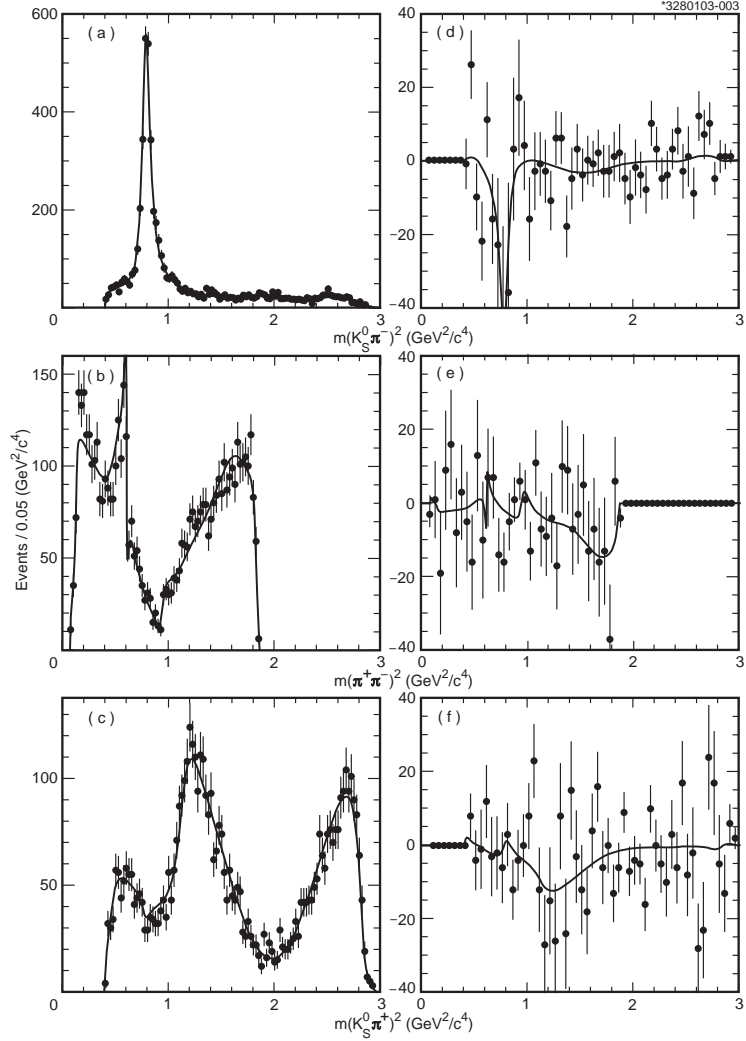,height=5.5in}
\end{center}
\caption{The projections of Dalitz fits on the $\d^0 \ra K_s \pi^+ \pi^-$
decays from the CLEO II.V data. The plots in the left column represent
the sum of the $\d^0$ and $\overline{\d}^0$ candidates. The plots in right
column represent the differences between the $\d^0$ and $\overline{\d}^0$
candidates. 
\label{fig:dalitz} }
\end{figure}

According to the Standard Model expectations, CP violation may happen
in the $\d^0 \ra K^0_S \pi^+ \pi^-$ decays at the $10^{-6}$ level due to
$K^0$--$\overline{K}^0$ mixing. Any observation of CP violation larger than
this level will
indicate new physics. Last year, we published a paper on this subject
using a newly developed Dalitz technique. Since this method measures decay
amplitudes rather than decay rates, the sensitivity to new physics is greatly
enhanced.  From the CLEO II.V data, we extracted the CP asymmetry in this
decay channel as
  $A_{CP} = -0.009 \pm 0.021 ^{ +0.010 + 0.013 } _{ -0.043 -0.037 }$.
which is interpreted as a non-observation of CP violation.~\cite{cpv}
Figure~\ref{fig:dalitz} shows the various projections of Dalitz fits on
the $\d^0 \ra K_s \pi^+ \pi^-$  decay candidates. The plots in the left
column represent the sum of the $\d^0$ and charge conjugate 
$\overline{\d}^0$ decays while the plots in right column represent the
differences between the $\d^0$ and $\overline{\d}^0$ decays.

\section{Summary and Future}
Based on the first 55.8 $\ipb$ data collected by CLEO-c at the $\psits$
resonance, we obtained a clean sample of $\dtomn$ decays and measured 
their decay constant as $\mathrm{f_{D^+}} = (202 \pm 41 \pm 17)\ \mev$. 
Using the same data set,
we produced the results on the exclusive branching fractions
of semileptonic decays of $\d^0$ and $\d^+$ mesons. The statistical power
of these results are already at the world best level, and two decay modes,
$\dotorhoen$ and $\dtowen$, were observed for the first time.  We 
are actively studying inclusive branching fractions of $\d$ mesons decays,
getting promising results.
Currently, CESR-c and CLEO-c are collecting more data at the $\psits$ resonance
with 12 wigglers. We have already collected $\approx$ 280 $\ipb$ at this energy
and we are planning to run on the $\d_s$ pair threshold area
and other interesting beam energies in future. 

\section*{Acknowledgments}
We gratefully acknowledge the effort of the CESR staff in providing us
with excellent luminosity and running conditions. This work was supported by
the National Science Foundation and the U.S. Department of Energy.

\end{document}